\documentstyle[preprint,eqsecnum,aps]{revtex}
\draft
\begin{document}
\newcommand{\be}{\begin{equation}}
\newcommand{\ee}{\end{equation}}

\title{{Tunneling in a bosonized Fermi liquid.}
{\it Release of \today}}
\author{L. S. Levitov$^{a,b}$ and A. V. Shytov$^b$}
\address{(a) Massachusetts Institute of Technology,
12-112,
77 Massachusetts Ave., Cambridge, MA 02139}
\address{(b) L. D. Landau Institute for Theoretical Physics, 2, Kosygin
st.,
Moscow, 117334, Russia}
\maketitle
\begin{abstract}
  We apply the bosonization technique to the problem of
tunneling in a Fermi liquid, and present a semiclassical theory
of tunneling rates. To test the method, we derive and evaluate
an expression for the tunneling current in the problem of two
parallel Luttinger liquids. Next we study tunneling between
parallel two dimensional Fermi liquids. In the absence of a
magnetic field, the conductance $I/V$ has a resonance peak at
zero bias. Our estimate for the peak broadening agrees with the
many-body theory results. In a magnetic field perpendicular to
the plane, we find a tunneling gap linear in the field, in
agreement with the experiment. In this regime the tunneling is
spatially coherent over a distance of the order of the cyclotron
radius. The coherence can be probed by a magnetic field parallel
to the plane, which gives an Aharonov-Bohm phase to the
tunneling amplitude. As a result, the current is an oscillatory
function of the field.
  \end{abstract}

\section{Introduction}
High  mobility  electron  gas in GaAs quantum wells forms a very
clean Fermi liquid. By varying electron density one  can  change the
effective   strength  of  the Coulomb  interaction,  and  study the
effects of
electron  correlation.  Of  special  interest  is   the
behaviour  of this system in a magnetic field, because it gives a
unique opportunity to study the properties of a Fermi liquid
in a  magnetic  field, with
virtually no disorder effects.

A new technique recently developed by Eisenstein
is a tunneling experiment
where two electron gases are confined in parallel wells separated by
a thin barrier through which electrons can tunnel from one well
to the other\cite{Gornik,Eisenstein'91,Eisenstein'91a}. Unlike a
conventional tunneling measurement, where electrons tunnel
through a link between two reservoirs or through a weak place
in a barrier, this system has a barrier that is very uniform.
Because of this, and also due to a very small amount of an
in-plane disorder, the momentum of a tunneling
electron is conserved. This gives two conservation laws for the
two components of momentum, which, together with the usual
energy conservation, highly restrict the phase space of possible
final states of the tunneling. As a result, the tunneling is of
a resonance character, and the tunneling current displays a
sharp peak. At close electron densities in the wells, the
tunneling conductance measured as function of the density
mismatch, has a peak of almost
Lorentzian shape, and of the width determined
by momentum relaxation time. This behaviour  is accounted for by
a free electron gas model\cite{ZhengMacDonald}.

A very interesting change in the tunneling current  is  observed
when  a  magnetic field is applied perpendicular to the plane of
the barrier. In the field, the resonance peak  is  shifted  away
from  zero bias by a certain amount, and a bias region is formed
in which tunneling conductivity is  almost  totally  suppressed.
The  physics of this ``tunneling gap'' can be understood in terms of
the work one has to do to pull an  electron  out  of  one  Fermi
liquid, and to inject it in the liquid on the other side of the
barrier\cite{Eisenstein'92,Eisenstein'94}. The role of the
magnetic field is that it
squeezes electron states and makes them localized in the
lateral dimension, which increases the energy of the Coulomb
interaction of an electron with its neighbors. The bias at which
there is enough energy to transfer an electron is bigger than
the Coulomb energy, and determines the gap and the
position of the peak.

The gap increases as a function of the magnetic field: linearly at
weak   fields,   when   there   are   many    occupied    Landau
levels\cite{Eisenstein'94,Turner},  and more slowly at higher
fields\cite{Eisenstein'92,Brown,controvercy}. Let  us  mention  that  a
similar  tunneling gap, linear in the field, was seen by Ashoori
in a tunneling capacitance experiment\cite{Ashoori}.

So far, theoretical work has dealt mostly with the gap in the high
field limit, where the problem has been treated using many
different
methods\cite{Halperin,HighFieldTheory,Efros,HighFieldExcitonic},
all arriving at the same estimate for the
gap: $\Delta\simeq e^2/\epsilon l$, where $\epsilon$ is the
dielectric constant, and $l$ is a characteristic interparticle
separation in the system. (In the Quantum
Hall regime, $l$ is of the order of magnetic length
$l_B=(hc/eB)^{1/2}$, and has a square root dependence on the
electron density.) In the low field limit, there is a
calculation by Aleiner, Baranger, and
Glazman\cite{Glazman,Aleiner}, which treats the problem
by using a hydrodynamical picture of an ideal conducting liquid
in a magnetic field, which is supposed to be valid at scales
much bigger than the cyclotron radius. In this work, the gap is
found to scale as a square of the magnetic field, which is at
odds with the experiment\cite{Eisenstein'94,Turner}. We think that the
reason for disagreement is that the scale relevant for the
problem is set by the screening length, which at a weak field is
much less than the cyclotron radius. Therefore, an appropriate
model should include Fermi liquid interaction effects, and thus the
 proper technique must be the Fermi liquid kinetic equation,
rather than the long wavelength hydrodynamics.

A conventional technique that deals with the problem of tunneling
in a Fermi system is the many-body perturbation theory, which
expresses the tunneling current in terms of one-particle
spectral weights, and then calculates the weights from the
Green's function perturbation series\cite{Mahan}. However, because the
interaction effects in this problem are in some sense always
strong, and also because in a weak field limit the problem
becomes semiclassical, a many-body perturbation theory
calculation, although possible, is not very illuminating.
Another option, less conventional in solid state theory, would
be to use a time-dependent Hartry-Fock mean field equation in
imaginary time, which is known in the theory of nuclear matter
to be capable of dealing with collective tunneling phenomena
such as nuclear fission\cite{Negele}.

In this paper we present an approach which better captures
 the semiclassical character of the problem at weak fields,
and treats it as a tunneling in a ``bosonized'' Fermi liquid. By
that we mean that the Fermi liquid is described in terms of a
Fermi surface fluctuating in space and in time. Of course,
such an approach is equivalent to the many-body
perturbation theory carried out in the long wavelength limit.
Our motivation for employing it, apart from interest in exploring a
new technique, is in its simplicity, and in the possibility of
establishing a clear relation with the phenomenological Fermi
liquid theory.

To clarify the last point let us recall that, when dealing with
a one particle tunneling problem, it is customary to consider an
equivalent classical problem and to evaluate the tunneling rate
exponent as a classical action on a path in imaginary time. The
Fermi liquid bosonization provides a natural extension of this
approach to the problem of tunneling into a Fermi system. One
can write an action which generates the Landau equation of the
phenomenological Fermi liquid theory, and then look for least
action paths in imaginary time. In such an approach one has
Landau quasiparticles, represented by deviations of the Fermi
surface from the spherical geometry, moving in imaginary time.
To determine the tunneling rate exponent, one has to find a
least action ``bounce'' solution of the Landau equation, which
describes a particle injected into the Fermi liquid, and then later
removed, and also the Fermi liquid response to the presence of
the particle.

Bosonization of a Fermi system is a well known method in the
one dimensional problem. It has been pointed out by Luther that
one can generalize the method  to higher dimensions\cite{Luther}, and
since recent work  by Haldane\cite{Haldane}  great
interest has been generated in this new
technique\cite{Fradkin,Houghton}.
We begin with a short review of the method; in dimension
one, and in many dimensions (Sec.~II). In Sec.~III, we
discuss how the tunneling problem can be treated in this formalism.

In Sec.~IV we turn to the specific problem of tunneling between
two parallel Fermi liquids. There are some unusual features of
this problem, from a technical viewpoint, that we discuss by
using an example of two parallel one-dimensional
Luttinger
liquids.
Due to the momentum conservation, the tunneling is
spatially coherent along the barrier. In the instanton technique
language, this is accounted for by summing over all points where
the particle enters the liquid, and where the particle leaves the
liquid.
 Such a summation
gives an ``entropical'' contribution, which properly accounts
for the conservation of momentum. By calculating the tunneling
current between Luttinger liquids, we demonstrate that such a
method agrees with the result obtained from the known exact
Green's functions. Another feature, explicit in the one
dimensional problem, is the role of momentum non-conserving
interaction. If the Luttinger liquid Hamiltonian has only a term
corresponding to scattering forward, there is a divergence in
the tunneling current at zero bias, accompanied by vanishing of
the current at any finite bias, which is expected from
what has been said above. Such a singular behaviour  can be
regularized either by introducing scattering by a disorder, or
by fluctuations of the barrier width.

The entropical contribution, as well as the regularization by a
momentum non-conserving scattering, are characteristic for the
multi-dimensional problem as well. In Sec.~V we generalize the
observations of Sec.~IV to the two dimensional problem, and also
demonstrate a relation to the classical orbits. It turns out
that, although we study dynamics in imaginary time, the spatial
structure of the instanton is characterized by classical orbits
found in real time, which are straight lines in the absence of a
magnetic field, and Larmor circles in a magnetic field.

Our new results,  all  contained  in  Secs.~V  and  VI,  can  be
summarized  as  follows.  In  the  absence  of the field, and at
matched electron densities in the wells, there is a peak of  the
tunneling  conductance  $I/V$  near zero bias. The peak width is
determined by momentum non-conserving scattering, and the  shape
of  the  peak is Lorentzian. The physics here is similar to that
of  the  peak  of  the  equilibrium  conductivity  $I/V_{I\to0}$
studied   as  a function   of   the   density   mismatch  in  the
wells\cite{ZhengMacDonald}.

At finite fields, which are much weaker than the Quantum Hall
fields, the peak is shifted to a finite voltage and a tunneling
gap is formed which scales linearly with the field. To compare
this result to the hydrodynamical
calculation\cite{Glazman}, we identify two terms
in our instanton solution, accounting for a quasiparticle
contribution to the gap, and for a collective Fermi liquid
response effect. The latter is identical to the hydrodynamical
theory result, however, the former is missing in it. The linear
field dependence agrees with the recent
experiment\cite{Eisenstein'94,Turner}. However, the dependence of the
gap on the electron density, predicted to be a square root, has
not been seen.

Also, we find that the tunneling is spatially coherent over a
large distance set by the cyclotron orbit radius, and propose to
probe this coherence by a magnetic field parallel to the
barrier. The advantage is that such a field does not affect the
in-plane motion, however, there is a dramatic effect on the
tunneling due to an Aharonov-Bohm phase in the tunneling
amplitude. The Aharonov-Bohm effect destroys coherence of
tunneling in the lateral dimension, and leads to oscillations of
the current taken as a function of the parallel field. In the
weak field limit these oscillations are given by a square of a
Bessel function.

\section{Bosonization formalism}
\subsection{One-dimensional bosonization.}

Bosonization of one dimensional Fermi systems\cite{Haldane-1D}
is a powerful tool for the investigation of properties of low
energy modes. It relies on the representation of the fermionic
operators in terms of bosonic ones. In the low energy limit the
current and the density of fermions become the only relevant
variables, which makes the bosonization possible. Here we
briefly review this method for the case of spinless fermions.

For making a relation to the higher-dimensional problem more
transparent,
instead of the density $\rho$ and the current $j$, it is
convenient to introduce the densities of right and left movers,
$\rho_{\pm}={1\over2}(\rho\pm j) = \psi_{\pm}^{\dagger}(x)
\psi_{\pm}(x)$,
which satisfy commutation relations:
\begin{eqnarray}
\lbrack \rho_{+}(x),\rho_{+}(x')\rbrack &=&
\lbrack \rho_{-}(x'),\rho_{-}(x)\rbrack =
 i\delta'(x-x')\ ;\nonumber\\
\label{commutator}
\lbrack \rho_{+}(x),\rho_{-}(x') \rbrack &=&0\ .
\end{eqnarray}
Then, one introduces the field operators
\be
\label{phi-field}
\phi_{\pm}(x) = \pm \sum_{p\neq 0} \frac{1}{p}
e^{-\alpha|p|/2 - ipx} \rho_{\pm}(p).
\ee
Here $\alpha$ is the ultraviolet cutoff set by the bandwidth.
Conversely, the representation for fermionic operators has the form:
\be
\label{boson-psi-1}
\psi_{\pm} = \frac{1}{\sqrt{2\pi\alpha}}
e^{\pm i k_Fx + i\phi_{\pm}(x)}\,.
\ee
The Hamiltonian of interacting fermions is
\begin{eqnarray}
H=\int dx\,
\Bigl\lbrack
&-&i v_F
\left(
\psi_{+}^{\dagger} \frac{\partial \psi_{+}}{\partial x}
-\psi_{-}^{\dagger} \frac{\partial \psi_{-}}{\partial x}
\right) \nonumber\\
&+&\frac{1}{2}\sum\limits_{a,b}{\cal U}_{ab}
\rho_a(x)\rho_b(x)
\Bigr\rbrack\ ,
\end{eqnarray}
where $a,b=+,-$, and $\cal U$ is a $2\times2$ matrix of
Fourier components of the interaction $U(x)$:
${\cal U}_{++}={\cal U}_{--}=U(0)$, and ${\cal U}_{+-}={\cal U}_{-+}=U(2k_F)$.
After the bosonization transformation, the Hamiltonian turns into
\be\label{hamiltonianD=1}
H = \frac{v_F}{4\pi} \int dx
\sum\limits_{a,b}
\left(\partial_x \phi_{a}\right)
\left(\delta_{ab} + F_{ab}\right)
\left(\partial_x \phi_{b}\right)\ ,
\ee
where $F_{ab}=\nu_{1D}\,{\cal U}_{ab}$. (Here $\nu_{1D}=1/2\pi v_F$ is
one half of the one-dimensional density of states.)
One can interpret the entries of the matrix $F$ as amplitudes of
forward ($F_{++}, F_{--}$) and
backward ($F_{+-}, F_{-+}$) scattering.
It will be seen later that the amplitudes $F_{ab}$
are 1D analogs of the Landau function
$F({\bf  n_1,n_2})$ of a Fermi liquid.

For considering tunneling in a Fermi liquid
we need an action corresponding to
Eq.(\ref{hamiltonianD=1}).
We derive it from the Hamiltonian
by following the method of Ref.~\onlinecite{Jackiw}.
 From Eq.(\ref{commutator}) we get the commutators of
$\phi_{\pm}(x)$, and use them to write the action:
\be S\!= -\frac{1}{4\pi} \int\!dx\!dt
\left(
\partial_t\phi_{+}\partial_x\phi_{+} -
\partial_t\phi_{-}\partial_x\phi_{-}
\right)
- \int\!H\!dt
\label{action-1}
\ee
Note that our Lagrangian describes two chiral field theories interacting
through backscattering.
The Lagrangian of a chiral field theory\cite{Jackiw} has the form:
$L_{ch} = \partial_x \phi \left( \partial_t +
\partial_x\right) \phi$. The solution of its
equation of motion is
\mbox{$\phi(x,t) = \phi_1(x-t) + \phi_2(t)$},
the second term $\phi_2$ being
a ``ghost,'' which does not contribute to
any observable
(e.g., to $\rho\sim\partial_x \phi$). One can say that the appearance of
the $\phi_2$
is the charge one pays for the locality of the Lagrangian.

\subsection{Multi-dimensional case}

It is well known that for a Fermi-system in any space
dimension, the relevant low energy excitations are
quasiparticles with momenta near the Fermi-surface. The Landau
theory of a Fermi liquid enables one to describe all low energy
physics in terms of a distribution of quasiparticles near the Fermi
surface, i. e., in terms of the fluctuations of the Fermi surface
geometry. Deformations of the Fermi surface are bosonic
excitations, and one may try to bosonize Fermi-liquid in $D>1$.
It has been done in
Refs.~\onlinecite{Luther,Haldane,Fradkin,Houghton}
 by a
procedure analogous to that used in dimension one. One
introduces normal displacement $u_{\bf n}({\bf r})$ of the Fermi
surface at the space point $\bf r$, in the direction ${\bf  n}$,
which means that ``local Fermi momentum''  is now ${\bf  n}(p_F
+u_{\bf n}({\bf r})$). Displacements $u_{\bf n}({\bf r})$
represent a multidimensional analog of $\rho_{\pm}$. Next, one
defines $\phi_{\bf n}({\bf r})$:
   \be
u_{\bf n}({\bf r}) = ({\bf  n\nabla}) \phi_{\bf n}({\bf r})\ ,
\ee
(cf. Eq.(\ref{phi-field})). Then, a canonical operator of an electron
can be written as
\be
\label{bosonic-psi-D}
\psi=\frac{1}{\sqrt{2\pi\alpha}}
\sum_{{\bf  n}} e^{ip_{F}{\bf  nr} + i \phi({\bf  n,r})}
\ee
(cf. Eq.(\ref{boson-psi-1})).
Commutation rules for $u_{\bf n}({\bf r})$,
\be
\label{commutator-D}
\lbrack u_{\bf n}({\bf r}), u_{\bf n'}({\bf r'})\rbrack =
i\delta_{{\bf  n,n'}}({\bf  n\nabla}) \delta({\bf  r} - {\bf  r'})\ ,
\ee
naturally lead to anticommuting $\psi$'s.
The Fermi-liquid Hamiltonian is
expressed in terms of $u_{\bf n}({\bf r})$ as
\be
H = \frac{\nu v_F^2}{2}
\int d{\bf  r'}d{\bf  n}\,
u_{{\bf  n}}({\bf  r}) \left(
(1+ \hat{F}) u\right)_{{\bf  n}}({\bf  r})\,.
\ee
Here $d{\bf  n}$ is a normalized Fermi-surface area (in two
dimensions, $d{\bf  n}=d\theta/2\pi$), and $\nu$ is the density
of states ($\nu_{2D} = m/\pi\hbar^2$). The integral operator $\hat{F}$
describes quasiparticle interaction in the Fermi liquid. For
simplicity, we assume the interaction to be of a density-density
type: $F= \nu U({\bf  r})$, where $U$ is the interaction
potential. It will be obvious to generalize the method to a more
general interaction $F({\bf  r}-{\bf  r'},{\bf  n}-{\bf  n}')$.

The action, by analogy with Eq.(\ref{action-1}), is
\be
S = -\frac{\nu v_F}{2} \int dt\,d{\bf  r} d{\bf  n}\,
({\bf  n\nabla}) \phi_{\bf  n}({\bf  r})\, \partial_t \phi_{\bf  n}({\bf
r})
- \int H dt\,.
\label{action-D-nomagn}
\ee
The equation of motion that follows from this action has the form
of the  Fermi liquid
kinetic equation:
\be
\left( \partial_t + v_F ({\bf  n\nabla})
\left(1+\hat{F}\right)\right)
u_{\bf n}({\bf r}) = 0\ .
\label{kin-equation-nomagn}
\ee

It is not entirely trivial to include a magnetic field in this picture
\cite{Haldane}.
It cannot be done by a na\"{\i}ve replacement
${\bf  \nabla}\to{\bf  \nabla}-{ie\over\hbar c}{\bf  A}$,
because in the Fermi surface description one has to
use the kinetic momentum $m{\bf  v}$, not the canonical one
${\bf  p} = m{\bf  v} + e{\bf  A}/c$. Hence, a gauge transformation
of $\phi$ is not just $\phi'_{\bf n}({\bf r})=\phi_{\bf n}({\bf r}) +
\chi({\bf  r})$.
To introduce a magnetic field, we consider the kinetic equation
\be
\label{kin-equation-mag}
\left(\partial_t + v_F D \left( 1 + \hat{F} \right) \right)
u_{\bf n}({\bf r}) = 0\,,
\ee
where
\be
\label{covariant}
D = {\bf  n}\cdot{\bf  \nabla}
- \frac{e}{c} {\bf  B}\cdot{\bf  n}\times{\bf  \nabla_{p}}\ .
\ee
Following Ref.~\onlinecite{Haldane}, we call this expression a
``covariant
derivative,'' since it takes into account the
curvature of the trajectory.
If $s$ is the length along the trajectory,
then the rate of change of $u_{\bf n}({\bf r})$ along it is given by
$\delta u/\delta s = Du$.

Now we redefine $\phi$ so that $u = D\phi$ and rewrite
the action to fit the  correct equation of motion.

\be
\label{action-D-mag}
S = - \frac{\nu v_F}{2} \int d{\bf  r}\,d{\bf  n}\,dt\,
D\phi\left( \partial_t + v_F (1 + \hat{F}) D\right) \phi\,.
\ee
This equation completes the description of the bosonized theory.

\section{Tunneling formalism}

Let us begin with some details on the system in which tunneling
experiments are done. Typically, it consists of two quantum
wells separated by a barrier of thickness $d\sim 8-20 nm$. The
wells' width $w$ is somewhat bigger: $w\sim 15-30 nm$. The
density of electrons in the wells is of the order of
$10^{11}cm^{-2}$. Usually electrons fill only the lowest energy
mode of transverse quantization. Because of this one can ignore
the motion perpendicular to the plane, and treat electron
dynamics as two-dimensional. The Coulomb screening length $r_s$
estimated from the density is of the order of $5-10 nm$.
However, due to the finite well width $w$, effective interaction
between electrons at distances shorter than $w$ is reduced
compared to the Coulomb potential, and this leads to the
increase of the screening length up to a few $w$'s. A magnetic
field's strength varies from $0.5 T$ to $5 T$ and higher,
corresponding to the filling factor varying from $\nu\approx10$
to $\nu\approx1$. For such fields, the magnetic length $l_B=(\hbar
B/ec)^{1/2}$ is of the order of $10-30 nm$, which is less or
comparable to the screening length.

In this paper we are going to deal with
cooperative effects that may strongly suppress charge tunneling.
During the  tunneling, the Fermi system has to
accomodate the presence of
a new electron,
and the action of this process will completely
control the tunneling rate. The single electron
tunneling, which is a part of this more complex process,
plays the role of an instant shake-up, followed by a  slower
relaxation in the many-body system.

In such a problem, one can use the tunneling Hamiltonian
formalism\cite{Mahan}. By neglecting complications such as a
finite well width, one can describe electrons in each well by a
two-dimensional Fermi liquid Hamiltonian. The wells are coupled
due to tunneling through the barrier, which is accounted for by
a tunneling term $H_T$ in the system Hamiltonian:
   \be
H_T = -  \int  t_0({\bf  r})\,
\psi^{+}_{R}({\bf  r})\, \psi_L ({\bf  r})\, d{\bf  r} + {\rm h.c.}
   \ee
Here $t_0({\bf  r})$ is a tunneling matrix element. In a real
system the barrier may be non-uniform, in which case $t_0$ acquires a
spatial dependence.

The tunneling current due to a voltage drop across the barrier $V$
can be found by a standard formalism\cite{Mahan}. First, one may
gauge the voltage away by transforming $\psi_R\rightarrow \psi_R
e^{ieVt}$. Then, one has to find the Matsubara transition amplitude,
   \be
\label{correlator}
K({\bf  r}, \tau) = \langle {\rm T}_\tau
\psi_R({\bf  r}, i\tau)  \psi_R(0,0)
\psi_L({\bf  r},i\tau) \psi_L(0,0)
\rangle\,.
\ee
To extract real-time information
from the Matsubara function $K$,
one continues $K$ to real time from the
upper and lower half-planes. (Following
Ref.~\onlinecite{Kadanoff},
we denote these functions as
 $K^{>(<)}({\bf  r},\tau)$).
The tunneling current $I$ is determined by
\be
I = 2\,{\rm Re}\,\int dt\,d{\bf  r}\,d{\bf  r'}\,
t_0({\bf  r}) t_0^{\ast}({\bf  r'})
\Phi_{{\bf  r} - {\bf  r'},t} e^{-ieVt}
\,,
\label{current}
\ee
 where $\Phi_{{\bf  r},t} = K^{>}({\bf  r},t) - K^{<}({\bf  r},t)$.

Now, let us derive an expression for the correlator in
Eq.(\ref{correlator}). We use
Eq.(\ref{bosonic-psi-D}) to
represent $\psi$'s. To evaluate the average, one has to write
down a path integral of the product of four exponentials with
the weight $\exp(-S)$. (From now on the time is imaginary.) This
Gaussian integral can be evaluated by a saddle-point method, so
that
   \be
\label{K-saddle}
K = \frac{1}{(2\pi\alpha)^2}
\sum_{{\bf  n_0,n_1}} e^{-S({\bf  n_0,n_1})}\,,
\ee
where $S({\bf  n_0,n_1})$ is the saddle-point action
\be
\label{Saddle-general}
S({\bf  n_0,n_1}) = \frac{2\pi}{\nu}
\langle
J D^{-1} \left(i\partial_t + v_F(1+\hat{F})D\right)^{-1} J
\rangle\,.
\ee
 $\langle\dots\rangle$  denotes $\int\dots d{\bf  n}d{\bf  r}dt$,
and
\be
J_{R(L)} = \pm
\left(
\delta({\bf  r})\delta(t)\delta_{{\bf  n,n_0}} -
\delta({\bf  r} - {\bf  R})\delta(t-\tau)\delta_{{\bf  n,n_1}}
\right)
\ee
Eq.(\ref{Saddle-general}) has a simple
interpretation in terms of the kinetic equation
(\ref{kin-equation-mag}).
Suppose an electron is instantly transferred
across the barrier at ${\bf  r}=0$, $t=0$, and then is
transferred back at ${\bf  r} = {\bf  R}$, $t=\tau$.
To incorporate the electron transfer, one must
add a source term to the
right-hand side of the kinetic equation (\ref{kin-equation-mag}):
\be
\label{kinetic-source}
\frac{\nu}{2\pi}
\left(i\partial_t + D (1+\hat{F}) \right) u  = iJ\,.
\ee
Then, one notes that Eq.(\ref{Saddle-general})
simply gives the action for this solution, describing the
Fermi liquid accomodation as an injection and removal of an electron.
Finally, according to Eq.(\ref{current}), one has to integrate
this action over ${\bf  R}$ and $\tau$, with the weight $e^{-ieV t}$.
This weight can
be interpreted as an additional term in the
action, which accounts for the change in the potential energy
during the tunneling time $t$ due to the voltage
source.

In this way, one may interpret
Eq.(\ref{Saddle-general})
as the action for a bouncing trajectory\cite{bounce},
and treat
Eq.(\ref{current}) as a sum of the amplitudes of all the different
bouncing trajectories. Quite often
this summation can be done simply by
taking the least action trajectory. However, in the case
of tunneling between parallel layers this conventional procedure
would lead to an incorrect result, because
there exists a great number of trajectories and their
``entropy'' contribution becomes essential.
As a technical remark, we will see that the summation over initial
and final points is responsible for momentum conservation
and can drastically change the answer.

In the next two sections, we apply
Eq.(\ref{Saddle-general})
to the problem of tunneling in a
one-dimensional Luttinger liquid and in a two-dimensional Fermi liquid.

\section{One dimensional problem}
\subsection{Clean system}
To illustrate how the formalism works,
we first consider tunneling in a one-dimensional
Fermi-system.
Let us take two parallel Luttinger liquids with a tunneling coupling
between them. For simplicity we include only interaction inside the
wire, but generalization is straightforward.
Then Eqs.(\ref{Saddle-general}) and (\ref{K-saddle})
 simplify to
(we introduce matrix notation, corresponding to left and
right movers) $K = \frac{1}{(2\pi\alpha)^2} \sum e^{-S_{\pm}}$
and
\be
\label{Saddle-Luttinger}
S = \pi \langle J
D^{-1} \left(i\partial_t + v_F (1 + \hat{F}) D\right)
J \rangle\,,
\ee
where
\be
J_{R(L)} = \pm
\left(
\delta(t)\delta(x) - \delta(t-\tau) \delta(x-X)
\right)
\left(
\begin{array}{c}
1\\
0 \\
\end{array}
\right)\,,
\ee
\be
D_{R(L)} = \partial_x
\left(
\begin{array}{cc}
1 & 0 \\
0& -1 \\
\end{array}
\right)\,
\quad
F_{R(L)} =
\left(
\begin{array}{cc}
F_1 &  F_2 \\
F_2 &  F_1 \\
\end{array}
\right)\,.
\ee
To find $S$ in Eq.(\ref{Saddle-Luttinger}), one introduces Fourier
transform
and carries out matrix inversion.
It gives
\be
S = 4 \int \frac{dkd\omega}{2\pi}
\frac{\omega - ikv_F(1+F_1)}{ik(\omega^2 + k^2 v^2)}
\sin^2\frac{\omega \tau + kX}{2}.
\ee
Here $v = v_F \sqrt{(1+F_1)^2 - F_2^2}$ is the velocity of excitations
in the Luttinger liquid\cite{Haldane-1D}.
After integration over $\omega$ and $k$, and summation over
left and right movers (and cutting off the divergent integral
at $1/\alpha$), one arrives at
\be
\label{K-Luttinger}
K(x,\tau) = \frac{\alpha^{2(\beta-2)}}{2\pi^2}
\frac{x^2 - v^2 \tau^2}{(x^2 + v^2 \tau^2)^{\beta}}
\,,
\ee
\be
\beta =
\sqrt{\frac{1+F_1-F_2}{1+F_1+F_2}} +
\sqrt{\frac{1+F_1+F_2}{1+F_1-F_2}}.
\ee
One can compare this expression to the product of two exact
Green's functions of the Luttinger liquid\cite{Luttinger-Green}.
Note that for a chiral liquid (i.e., with no backscattering,
$F_2 = 0$) one has $\beta=2$, and only Fermi-velocity is
renormalized. Note also that $\beta$ does not depend on the sign
of $F_1$ and $F_2$ to the first nonvanishing order of the
perturbation theory in $F_1$ and $F_2$. This happens because of
an approximate ``particle-hole symmetry": if the interaction is
attractive, the work should be done to remove a particle, while
for a repulsive interaction the same work is required to inject
a particle.

To get the tunneling current, one has to compute the integral in
Eq.(\ref{current}). On dimensional grounds, one expects the current to
behave as a power law: $I\sim V^{2(\beta-2)}$. Explicit calculation
(see Appendix A)
gives
\begin{eqnarray} \label{I-Luttinger}
I(V) &=& C(\beta)\,\frac{e|t_0|^2}{v}\,
\left(\frac{eV\alpha}{v} \right)^{2(\beta-2)}\,,\\
C(\beta) &=& \frac{1}{\sqrt{\pi}}\,
\frac{(\beta-2)\, \Gamma
\left(
\beta- \frac{3}{2}
\right)}{\Gamma(\beta)\, \Gamma(2\beta -3)}\ ,
\end{eqnarray}
where $\Gamma(\beta)$ is the $\Gamma$--function. However, in an
apparent contradiction to the dimensional estimate, for the
simplest case of free fermions, Eq.(\ref{I-Luttinger}) gives
$\beta = 2$, $C(\beta) = 0$, and $I(V)=0$. The same cancellation
of current also occurs in the absence of backscattering.

This cancellation has a clear physical reason. It has been
pointed out\cite{Eisenstein'91}
that, in a gas of free fermions the
energy and momentum conservation
prohibit tunneling
at any nonzero voltage.
This argument works even for an interacting system
in the absence of backscattering. However,
it fails when the backscattering is present,
because of the possibility
of an electron scattering to another
branch of the   spectrum, with the $2p_F$ change in momentum.

Therefore,
to get a finite tunneling current for a system with no backscattering,
one needs to introduce some mechanism which changes
the momentum of the tunneling electron (i.e.,
breaks
translation invariance).
There are two mechanisms in real systems:\\
(i) scattering by a disorder;\\
(ii) spatially fluctuating
barrier width;\\ or, more   formally, a spatial dependence
of the tunneling matrix element.
If the barrier is not uniform, the tunneling occurs
preferentially where it is low or narrow. Below
we consider both mechanisms.

\subsection{Scattering by a disorder}

We consider here only weak disorder, so that electron motion
can be treated as ballistic. This is the case when   the characteristic
time of motion is much smaller than the scattering time.
In this limit one may take scattering into account semiclassically.
Namely, since the probability for the electron
to move by a distance $s$ without
being scattered is equal to $\exp(-s/l_s)$
($l_s$ is scattering length),
we can simply multiply the amplitude $K$ by this factor.
After such a ``regularization
by scattering,'' one gets
\be\label{Phi-disorder}
\Phi_r(x,t) = \Phi(x,t) \exp(-|x|/l_s).
\ee
One can compare this result to the perturbation theory
expression for the Green's function in a random potential.
In a ladder approximation~\cite{AGD}, the Green's function average
 over disorder
differs from that in a clean system by the factor
$\exp(-r/2l_s)$.

After substituting Eq.(\ref{Phi-disorder}) in
Eq.(\ref{current}), and
integrating over $x$ and $t$, one arrives at
\be
\label{I-1D-free-scattering}
I(V) = \frac{2 e |t_0|^2}{\pi l_s}
\frac{eV}{(eV)^2 + (\hbar/\tau_s)^2}
\, ,
\ee
where $\tau_s=l_s/v_F$. So, the effect of elastic scattering is
that the conductance $I/V$ becomes a Lorentzian function.

Note that at small   $V$
Eq.(\ref{I-1D-free-scattering})
 gives Ohm's law, which
differs from the power law
(\ref{I-Luttinger}).
This occurs because of the presence of
an additional dimensional parameter $l_s$.

Note also that the tunneling conductance in the Ohmic regime is
proportional to $l_s$. Such a result is unusual for a tunneling
problem. For example, for tunneling in a point contact, the
$I$--$V$ dependence is also Ohmic, but the tunneling conductance
is expressed solely through the density of states, and the mean
free scattering path $l_s$ does not enter. Here, the appearance of
$l_s$  reflects an important feature of tunneling between
two layers: together, the energy and momentum conservation
prohibit tunneling at any nonzero voltage, and  due only to
elastic scattering does the tunneling become possible. One also may
note that the behaviour  of the tunneling current as $l_s
\rightarrow \infty$ is extremely singular.

To summarize, the disorder leads to the broadening of the
resonance tunneling peak at $V=0$ into a Lorentzian peak of the
width $\hbar/\tau_s$.

\subsection{A non-uniform barrier}

Since the probability to tunnel depends on the width of the
barrier exponentially, the tunneling through a barrier of a
fluctuating thickness occurs at the points where the barrier is
the thinnest. Near such points one can use the following model
for the tunneling matrix element:
  \be
\label{nonuniform-tunnelling}
|t_0(x)| \approx \frac{A}{\sqrt{2\pi\sigma^2}}
\exp\left(- \frac{x^2}{2\sigma^2}\right)\,.
\ee
Here $\sigma$ is the lateral scale of the barrier fluctuation.
Since the motion of an electron in this model
is always ballistic, one may assume an
arbitrary voltage.
By doing the integral in Eq.(\ref{current}), one gets
\be
\label{I-1D-free-barrier}
I(V) = \frac{2}{\pi} \frac{A^2}{v_F^2} eV
\exp\left(
 -\left(\frac{eV\sigma}{\hbar v_F} \right)^2\right)\ .
\ee
Note that, compared to the previous case,
the suppression of tunneling at
large $V$ is stronger. Also,
in the limit $\sigma\rightarrow 0$, the
barrier
turns into a  point contact, and one recovers from
Eq.(\ref{I-1D-free-barrier}) the usual Ohm's law.

To summarize, the semiclassical treatment of tunneling through
an extended barrier is different from that in a point contact.
We will encounter a similar situation in the discussion of a two
dimensional problem, so, to facilitate a comparison, let us list
the main things we learn from the one dimensional tunneling
problem:\\
(i) The summation over initial and final points
is an essential feature of the formalism, and
can change the answer drastically;\\
(ii) The tunneling current for free
particles has to be regularized by
breaking translational invariance; and\\
(iii) The answer is sensitive
to the regularization scheme,
which has to be chosen on physical grounds.

\section{Two dimensional problem: no interaction}

It turns out that for 2D  electrons  in  a  magnetic  field  the
technically  difficult  part  is  to carry out the semiclassical
calculation for free particles. After that, taking into  account
the interaction will be straightforward.

\subsection{Relation of the least action paths to classical orbits}

Now, we will consider free electrons, $\hat F=0$,  and  evaluate
the  action  (\ref{Saddle-general}).
It  is worthwhile to mention
that because of the semiclassical character of the problem,  the
calculation  will naturally lead to a relation with the cyclotron
orbits of the classical problem.
To deal with  the  operator  in
(\ref{Saddle-general}),
it is useful to use Fourier-transform
 \be
\phi = \sum_{k,\omega}
\phi_{k\omega} e^{-i\omega t + i{\bf k}{\bf x}}
\ee
and to  look for a solution to the equation
\be
(i\partial_t + D)
G(\theta,\theta') = \delta(\theta - \theta')\,,
\ee
where $\theta$ is polar angle: ${\bf n}=\cos\theta{\bf i}+\sin\theta{\bf j}$.
The solution
is  given by
\be
\label{inverted_operator}
G(\theta,\theta') = \sum_{n}
\frac{e^{in(\theta - \theta')
+ i{\bf k}\times({\bf n} - {\bf n'})
R_c}}{2\pi(\omega + i n \omega_c)}
\ee
where $\omega_c = eB/mc$ is
Larmor frequency and $R_c = v_F/\omega_c$ is
Larmor radius.

Now, it is straightforward to evaluate the saddle-point action:
\begin{eqnarray}
S_{free} = - \frac{4}{\nu} \int \frac{d^2k d\omega}{(2\pi)^3}
\sum_n
\frac{1}{in\omega_c( \omega+ in\omega_c)}
\times\nonumber\\
\sin^2\frac{1}{2}(n(\theta_0\!-\!\theta_1)+
{\bf k}
\left(\hat{{\bf  z}}\!\times\!({\bf n_0}\!-\!{\bf n_1})R_c\!+\!
{\bf R}\right)
 - i\omega\tau)
\ .
\label{saddle-2D}
\end{eqnarray}
At this point we encounter a difficulty: the expression
(\ref{saddle-2D}) is
formally divergent. For electrons without a magnetic field
it has been noted\cite{Luther} that the action is finite
only when there exists a classical trajectory with
given momentum going from $0$
to ${\bf  R}$. If this is not the case,
the transition amplitude becomes zero.
Now, we will see that the same situation occurs in a magnetic field.

To begin with, one
divergence comes from the
$0-$th term of the sum in Eq.(\ref{saddle-2D}).
This divergence will be eliminated, if we impose the
condition
   \be
{\bf \hat{z}} \times ({\bf n_0} - {\bf n_1}) R_c +
{\bf R} = 0\,.
\label{coincidence}
   \ee
In the classical language,
Eq.(\ref{coincidence}) means that the initial and final points
belong to the same cyclotron orbit.
However, this does not
solve the whole problem, because now
we have a diverging ${\bf  k}$--integration, which gives
\be
\label{delta-function}
\delta(\hat{{\bf  z}} \times ({\bf  n_0} - {\bf  n_1}) R_c
+ R ) =
\delta(0)\ .
\ee
To regularize this $\delta$--function, let us note
that Eq.(\ref{coincidence})
 has semiclassical accuracy,
because it implies a definite trajectory of the
particle. However, in a quantum problem there is a finite thickness to
anything,
and thus the trajectory is washed out a bit,
so that the   $\delta$--function effectively has a finite width.
To determine this width, let us recall that,
in  a  semiclassical picture, different states of a particle can
be associated with the cells in the phase space  of  the  volume
$(2\pi\hbar)^d$, where $d$ is the space dimension.
 From this, a regularization of the
phase space $\delta$--function follows:
\be
\label{regularization-xp}
\delta_p(0) \delta_x(0) = \frac{1}{2\pi\hbar}\ ,
\ee
where the subscripts denote the
arguments of the  $\delta$--functions.
To regularize the   $\delta$--function in
Eq.(\ref{delta-function}),
we rewrite it in terms of
the   coordinate and momentum:
\be
\label{regularization}
\delta(x)\delta\left(\frac{c p_x}{eB} \right) =
\frac{eB}{c} \delta(x)\delta(p_x) = \frac{eB}{2\pi\hbar c}
\,.
\ee
Here, we drop higher order terms in the $\delta-$function argument,
since we are interested in
dimension only.

Therefore, we find that:\\
(i) The action is finite only if there exists a corresponding
classical trajectory;\\
(ii) The regularization of the action can be done
semiclassically.

Let us mention a similarity of our regularization to the
regularization of $\delta^{(4)}(p)$ in the scattering amplitude
in quantum field theory\cite{QFT}, done by a finite volume.
(A finite volume problem with periodic boundary conditions,
leads to a lattice in the momentum space, and a lattice
$\delta$--function is finite.)

Now, the regularized action is finite:
\be \label{action-free}
S_{free} = - \sum_n \int \frac{d\omega}{\pi}
 \frac{ 1 - \cos\left(n(\theta_0 - \theta_1) - \omega\tau\right) }
{in(\omega + in\omega_c)}\ .
\ee
The integral over $\omega$ can be done by the residue
method, and the sum can be calculated by using the
identity
\be
\label{identity}
\sum_{n=1}^{\infty} \frac{e^{-nz}}{n} =
\log \frac{1}{e^z - 1}\,,
\qquad \left({\rm Re}\,z > 0\right)\ ,
\ee
and cutting it off for $z=0$ divergent part at
$n_{max} = \omega_c/v_F\alpha$ (Recall that
$\alpha$ is a short time cutoff of the order of
inverse Fermi-energy or bandwidth).
Finally, the answer reads
  \be
K({\bf R},\tau) = \sum_{{\bf n}_0, {\bf n}_1}
\left(\frac{\omega_c}{4\pi v_F
\sin^2(i\omega_c \tau + \theta_0 - \theta_1)/2} \right)^2
\ ,
\label{Kfree}
  \ee
where the sum is restricted to the points ${\bf n}_0$, ${\bf
n}_1$, ${\bf R}$ of the cyclotron orbit given by
Eq.(\ref{coincidence}).

 From Eq.(\ref{inverted_operator}),
a simple intuitive picture of
tunneling  follows.
If one looks for a solution to
the kinetic equation (\ref{kinetic-source}), one
finds that it corresponds to the quasiparticle
uniformly distributed along a cyclotron orbit
at $\omega_c \tau \gg 1$. Clearly, it simply
means
that in a translationally invariant system
only the tunneling between
cyclotron orbits with the same centers is allowed, and thus one
can think of a tunneling electron spread over a cyclotron orbit
and hopping to an identical orbit in the other plane.

\subsection{Semiclassical Jacobian}
The sum in Eq.(\ref{Kfree}) still requires some care, because of the
semiclassical accuracy of Eq.(\ref{coincidence}).
Na\"{\i}vely, one would simply take as a final answer
the expression (\ref{Kfree})
for ${\bf  n_0}$ and ${\bf  n_1}$ satisfying
Eq.(\ref{coincidence}), but this turns out to be incorrect.
It is known\cite{Gutzwiller} that a
semiclassical propagator contains as
a prefactor the Jacobian
$\partial p_i/\partial q_j$, describing
the divergence
rate of two trajectories going
from the same point with different momenta.
The role of this Jacobian is analogous to the role
of the prefactor
of the semiclassical
wave function in dimension one:
$\psi\sim p^{-1/2}\exp(iS)$.

Below, we present an "intuitive" way to get the Jacobian from the
summation over ${\bf  n}_0$, ${\bf  n}_1$ in
Eq.(\ref{Kfree}).

To carry out the summation, let
us consider the initial wave packet of the
width $\delta x$ in a direction
transverse to the trajectory.
This packet has, therefore, a transversal momentum
distribution of the width $\delta p = 2\pi\hbar/\delta x$.
For  free motion in a magnetic field, the corresponding uncertainty
of the final point of the trajectory
in the transverse direction is
$\delta x' = \delta p /(p_F R_c\sin(\theta_1 - \theta_0))$.
To have  non-zero probability of finding
the particle at ${\bf  R}$,  $\delta x'\le\delta x$ is required.
 By putting all together,
one derives the semiclassical interval of
angles over which the sum should be taken:
\be
\delta \theta = \frac{\delta p}{p_F} =
\sqrt{\frac{2\pi}{p_F R_c \sin(\theta_1 - \theta_0) }}
\ee
 From this, one can get  the  Green's function.
By adding the amplitudes for $({\bf  n_0,n_1})$
and $({\bf  -n_1,-n_0})$, one has
\begin{eqnarray}
\label{K-Jacobian}
K({\bf  r},\tau) =
\left(
\frac{\delta \theta}{2\pi}
\right)^2
\left(
\frac{\omega_c}{4\pi v_F}
\right)^2 \times \\
\left\lbrack
\left(
\frac{1}{\sin\left(i\omega_c\tau - \theta({\bf  r})\right)/2}
\right)^2
+ \left(\theta \rightarrow -\theta\right)
\right\rbrack\,,
\nonumber
\end{eqnarray}
where
$\theta({\bf r}) = \theta_0 - \theta_1  = 2 \sin^{-1} r/2R_c$.
Now, for verification,
one can check Eq.(\ref{K-Jacobian}) with the
semiclassical expansion of the exact Green's function.

Finally, let us remark that the inability of our semiclassical
method to automatically generate a correct Jacobian is a general
property of a theory with linearized spectrum. The reason is
that, in such a theory wavepackets do not spread, due to a linear
dispersion relation, and thus the Jacobian must equal one. For
example, in a one dimensional problem with a linear spectrum, in
the semiclassical expression for the wavefunction the prefactor
$p^{-1/2}$ is absent. Taken literally, this leads to various
anomalies, e.g., to the absence of a linear response to an
electric field. In the one dimensional problem, this drawback is
eliminated by incorporating a chiral anomaly into the theory.
The bosonized theory reflects the anomaly as the Schwinger term
in the right hand side of Eq.(\ref{commutator}). The multidimensional
problem, however, does not incorporate it in a proper way. More
exactly, only the ``longitudinal" part of it is present in
Eq.(\ref{commutator-D}), because the Fermi surface is assumed to be
locally flat, while its actual non-zero curvature is
 responsible
for the ``transverse" part that leads to the Jacobian obtained above.
Also, in our opinion, the same physical reason applies to all the
divergences discussed in the previous subsection.

\subsection{Regularization by scattering}

We are going to introduce elastic scattering
to regularize the expression for the tunneling
current, similar to how it was done in one dimension. In a higher
dimension,
it turns out to be a more
complicated procedure, because  of the uncertainty
of the particle path going to a given final
point ${\bf  r}$. In one dimension, the path length would be simply
$|r|$, however, in two dimensions,
since the motion in the
magnetic field is periodic,
the particle can make $0,1,2,\dots$ complete circles
before arriving at ${\bf  r}$.
One can resolve this difficulty, by
using the identity
\be
\label{identity-more}
\frac{1}{\sin^2 z} = \sum_{l=-\infty}^{\infty}
\frac{1}{(z-\pi l)^2}\ ,
\ee
and expanding
the amplitude $K$ in the following series
form:
\begin{eqnarray}
\label{K-expanded}
K =
\left(
\frac{\delta \theta}{2\pi} \frac{\omega_c}{2\pi v_F}
\right)^2\times\\
\sum_{l}
\left(
 \frac{1}{(i\omega_c\tau - \theta - 2\pi l)^2} +
\left(\theta\rightarrow -\theta\right)
\right)\nonumber\,.
\end{eqnarray}
By comparing it to the exact one-dimensional
Green's function, one can see
that the $l-$th term in this expression can
be interpreted as an amplitude
of $l$ rotations before arriving at ${\bf  r}$.
In this way, the path for each term is
$s_l = R_c|\theta + 2\pi l|$.
Therefore, due to  elastic scattering, each term is reduced
by $\exp(-s_l/l_s)$.
In this way, the regularized amplitude takes the form:
\begin{eqnarray}
\label{K-regularized}
K_{r} =
\left(
\frac{\delta \theta}{2\pi}
\frac{\omega_c}{2\pi v_F}
\right)^2 \times\\
\sum_{l}
\left(
\frac{e^{-|\theta + 2\pi l|R_c/l_s}}{(i\omega_c \tau
- \theta - 2\pi l)^2}
+ \left( \theta \rightarrow -\theta\right)
\right)\nonumber
\end{eqnarray}
The scattering acts as a cutoff to this sum. Effectively, it
eliminates the contributions of the trajectories which wind
around the orbit so many times that their length exceeds the
scattering length $l_s$.

In order to be able to apply this method, one has to have a
ballistic situation, \mbox{$R_c\gg l_s$}. If this is not the case,
the use of a collisionless Fermi-liquid equation to determine the
motion is incorrect. Therefore, this method  enables one to
deal only with a classically strong magnetic field.

\subsection{Calculation of tunneling current}
Finite width of the experimentally observed peak has been attributed to
the fluctuations of barrier width\cite{Turner}. Here we calculate the effect
on tunneling current due an in-plane elastic scattering. We
expect that the broadening of the peak due to momentum non-conserving
scattering (no matter what origin)
will be similar to that found in Sec.IV for the one-dimensional problem.

To get the tunneling current from
Eqs.(\ref{current}) and (\ref{K-regularized}),
we continue $K$ to real time:
\begin{eqnarray}
K^{>}({\bf  r},t) - K^{<}({\bf  r},t) =
\frac{i}{p_F R_c \sin\theta({\bf r})}
\left(\frac{\omega_c}{2\pi v_F}\right)^2
\nonumber\\
\sum_{l}
\delta'(\theta - \omega_c t + 2\pi l)
e^{-|\theta + 2\pi l|R_c/l_s}
\label{K-continued}
\end{eqnarray}
Now one plugs this expression into
Eq.(\ref{current}) and integrates it.
Here it is useful to reduce the integration over
${\bf  r}$ to the integration over $\theta$:
$d^2 {\bf  r} = 2\pi r dr = 2\pi R_c \sin\theta d\theta$,
since $2 R_c \sin\theta/2  = r$.
The final answer reads
\be
I_{free}(V) = \frac{2}{\pi} \frac{e|t_0|^2}{l_s p_F}
\frac{eV}{(eV)^2 +(\hbar/\tau_s)^2}\ ,
\label{I-free}
\ee
where $\tau_s=l_s/v_F$. Note that
this expression is identical, up to a prefactor,
to the result for free electrons in the dimension one, given by
Eq.(\ref{I-1D-free-scattering}), and the same discussion applies here.

\section{Interacting electrons in a magnetic field}
\subsection{Effective interaction}
In this subsection we treat $\hat{F}$ in
Eq.(\ref{Saddle-general}). We
start with an expansion of Eq.(\ref{Saddle-general}) in powers of $F$,
and rewrite it as a sum of two terms:
  \be
\label{two-terms}
\frac{1}{D\left(\omega\!+\!D(1\!+\!\hat{F})\!\right)}  =
\frac{1}{D\left(\omega\!+\!D\right)} +
\frac{1}{\omega\!+\!D}\hat{\cal{F}}\frac{1}{\omega\!+\!D},
\ee
where
\be
\label{ladder}
\hat{\cal{F}} = \hat{F} +
\hat{F} \frac{D}{\omega+D}\hat{F} + \cdots =
\hat{F} + \hat{F}\frac{D}{\omega + D} \hat{\cal{F}}
\ee
The first term on the right hand side of
Eq.(\ref{two-terms})
yields the free particle action (\ref{action-free}) studied
above. The ``effective'' interaction $\cal{F}$ in the second
term is defined by the Fermi liquid ladder, given by
Eq.(\ref{ladder}). As usual, Eq.(\ref{ladder})
can be treated as a
Dyson equation.
It describes the screening of the
interaction  $F$ by the electron gas.

To solve Eq.(\ref{ladder}), we rewrite it as
\be
\hat{\cal{F}}(1 + \hat{F}) = \hat{F} +
\hat{F} \frac{\omega}{\omega + D}
\hat{\cal{F}}
\ee
and use Eq.(\ref{inverted_operator})
for $(\omega + D)^{-1}$.
If $F$ does not depend on
angles (like for a density-density type interaction), a
convolution gives
\be
\label{ladder-1}
(1 + F_k ) {\cal{F}}_{k\omega} = F_k
\left(
     1 +
\omega \sum_n \frac{J_{n}^2(kR_c)}{\omega + in\omega_c}
                       {\cal{F}}_{k\omega}
\right)
\ee
Here $J_n$ are Bessel functions.

Let us solve Eq.(\ref{ladder-1}) for small $\omega\ll\omega_c$. The range of
$\omega$ is set by relevant voltage, which in its turn is
determined by the effective interaction. We will see below that the
latter is small compared to $\hbar\omega_c$, and this
implies that the voltage $V$ is
small:  \mbox{$eV \ll \hbar \omega_c$}.

One has to take the limit $\omega \rightarrow 0$ carefully,
because of the singular nature of the operator $D^{-1}$.
If one simply sets $\omega = 0$ in
Eq.(\ref{ladder}), one incorrectly  gets ${\cal F}=F/(1+F)$.
However, the presence of a finite $\omega$ changes
the result.
One may note that the limits $n\rightarrow 0$ and
$\omega\rightarrow 0$ in the  $0-$th term of the
sum in Eq.(\ref{ladder-1}) do not commute.
The origin of such  behaviour  lies in the existence
of an infinite number of stationary solutions
to the   kinetic equation Eq.(\ref{kin-equation-mag}).
Indeed, if one neglects by
the time derivative in
Eq.(\ref{kin-equation-mag}), one
finds that there exist
an infinite number of
solutions
to the equation $D\tilde{u} = 0$, where
$\tilde{u} = (1 + \hat{F})u$.
These solutions are axially symmetric charge distributions,
uniformly rotating in the magnetic field.
But any finite frequency, no matter how small,
excludes such solutions.

Thus, for $\omega \ll \omega_c$, Eq.(\ref{ladder-1}) turns into
\be
\label{F-renormalized}
{\cal{F}}_{k} = \frac{F_k}{1 + F_k\left( 1 - J_0^2(kR_c)\right)}.
\ee
By using asymptotical expressions for the Bessel function, one
finds that for $kR_c \gg 1$ the interaction is screened, like in
a metal, while for $kR_c \ll 1$ the behaviour  is
like in a  dielectric.
 This reflects localization of electrons by a magnetic field.
In reaction to a small external electric field, the cyclotron
orbits are shifted by a small distance along the field,
and also start drifting perpendicular to the field. Effectively, this
leads to a dielectric response to an external electric field.

\subsection{Tunneling gap}
The interaction
contribution  to the saddle point
action is $ J \left(\omega-D\right)^{-1}
{\cal{F}}
\left(\omega- D\right)^{-1} J$.
It is straightforward to find  that
\be
S_{int} - S_{free} = \frac{2}{\pi m } \int \frac{d^2 k \,
d\omega}{\omega^2}
{\cal{F}}_k J_0^2(kR_c) \sin^2 \frac{\omega \tau}{2}
\ee
Integrating over $\omega$, one finds
$
S_{int} - S_{free} = |\tau| \Delta
$, where
\be
\label{Gap}
\Delta  =
 \int
\frac{J_0^2(kR_c) F_k}{m\left(1 +
F_k\left(1-J_0^2(kR_c)\right)\right)}\,
k\,dk .
\ee
Since the action is linear in time, the quantity $\delta$  is the energy
of a static configuration, which eventually forms at $t\gg
1/\omega_c$ after tunneling.

By analytic continuation of $t$, this contribution to the action
turns $K^{>(<)}$ into
\mbox{$K^{>(<)}_{free}e^{\pm i\Delta t}$}.
Now, it is useful to integrate in
Eq.(\ref{current}),
first over $t$, and then over $d{\bf  r}$.
The integral over $t$ can be easily evaluated by the
residue method, since the time integration contour can be closed.

It turns out that, because of the position of the poles
of $K^{>(<)}$, the contribution of
$K^{>}$ is zero unless $eV > \Delta$, while the
contribution of $K^{<}$ is zero unless $eV < -\Delta$.
This leads to the result:
    \be
\label{MainResult}
I =
\left\{
\begin{array}{rcl}
I_{free}(V - \Delta) &{\rm at}& V>\Delta\ ;\\
0 &{\rm at}& -\Delta<V<\Delta\ ;\\
I_{free}(V + \Delta) &{\rm at}& V<-\Delta
\end{array}
\right.\ ,
\ee
where $I_{free}$ is given by Eq.(\ref{I-free}).

Hence, the $dI/dV$ curve has peaks at $eV = \pm \Delta$, and a
gap of the width $2\Delta$ between them. The applicability of
Eq.(\ref{MainResult}) is restricted to the vicinity of the
peaks: it correctly gives the width of the gap and the current
near the peak, but it does not hold deep inside the gap, near
$V=0$, because our simple approach to scattering breaks down at
$t \gg \tau_s$. In this limit, the character of charge spreading
changes from ballistic to diffusive, and one has a Coulomb
anomaly problem in a diffusive system (e.g., see
Ref.~\onlinecite{LS1}).

\subsection{Qualitative discussion}
One can make a simple qualitative picture of the tunneling on
the basis of Eqs.(\ref{two-terms}) and
(\ref{F-renormalized}). Note that the
first term in Eq.(\ref{two-terms}) yields a solution of the
non-interacting Fermi liquid equation, which is is localized on
the cyclotron orbit in space, and in the limit $\omega_c\tau \gg
1$ possesses an axial symmetry. This solution has a
$\delta-$function singularity in the momentum space,
because at each point of the orbit the particle momentum is
tangential to it.

In the  presence of interaction,
the solution has an extra term, which can be obtained by
applying $\hat{\cal{F}}$ to the solution for the non-interacting
problem.
First, since the operator $\hat{\cal{F}}$
integrates over all angles in the momentum space,
the resulting contribution to $u_{\bf n}({\bf r})$
is a smooth function in the momentum space. Also,
one can see that it has a spatial scale
$a_{0}\sim \nu e^2 R_c^2 \gg R_c$
(at this scale the denominator
in Eq.(\ref{F-renormalized})
becomes the order of $1$).
Therefore, the two terms of the total solution are:\\
(i) a quasiparticle like $\delta-$function
singularity, localized
on the classical orbit; and\\
(ii) a  smooth background.\\
In Ref.~\onlinecite{Glazman} the second contribution was derived
hydrodynamically, while the first one was missed, since it
corresponds to the ballistic motion of one particle and cannot
be treated hydrodynamically. In fact, we think that the
applicability of hydrodynamics for treating this problem is
questionable, since one deals here with the collisionless limit.
Nevertheless, our consideration justifies the picture of
Ref.~\onlinecite{Glazman}, and corrects it only in the region
$r\sim R_c$ where hydrodynamics is not applicable. This region
corresponds to the core of the ``vortex'' of Aleiner, Baranger, and
Glazman.
Nevertheless, it turns out that the contribution
of the core to the gap becomes essential in the
two-layer
system, when the electron at one plane interacts
with a hole on
the other plane.
In the one-layer system, where one should
substract self-interaction from Eq.(\ref{Gap}),
the result of Ref.~\onlinecite{Glazman} gives
the correct value of the gap.

The electrostatic energy of the core is determined by the
inverse cyclotron radius (see Ref.~\onlinecite{LS2}), while the
energy of the smooth background is proportional to $a_0^{-1}$,
and thus is much smaller.

Let us  analytically evaluate the
contributions to $\Delta$ from 2 regions
in the $k-$space, by using the asymptotics of
$J_0(kR_c)$
in the regions
$kR_c \ll 1$ and $kR_c \gg 1$.
We take the interaction $F$ as the
difference  of the   in-plane and
interplane Coulomb interaction,
which accounts for the opposite sign image in the other plane.
The Fourier transform of this effective Coulomb interaction,
with an image contribution from the other plane,
is $U_k = 2\pi e^2 (1-e^{-kd})/k$,
and $F_k = \nu U_{k}$.
Assuming that  $d\ll R_c$, $d\gg r_s$ and
$l_B \ll r_s$ ($r_s = (2\pi\nu e^2)^{-1}$ is the Debye
screening length,
$l_B = \sqrt{\hbar c / eB}$ is magnetic length),
one has
\begin{eqnarray}
\label{ABG}
\Delta_{kR_c \ll 1 } &=&
\frac{\omega_c^2}{m v_F^2} \ln\frac{d}{r_s}\\
\label{capacitor}
\Delta_{kR_c \gg 1}  &=&
\frac{e^2 \omega_c}{\pi v_F} \ln \frac{r_s}{l_B},
\end{eqnarray}
The magnetic length  was
chosen as an ultraviolet cutoff, since at this distance the
semiclassical approximation breaks down.

The contribution (\ref{ABG}) coincides with that found by a
hydrodynamical approach in Ref.~\onlinecite{Glazman}, the only
difference being the effect of screening by another plane.
{}From
Eq.(\ref{capacitor}) one can see that the region
$k \gg 1/r_c$ does give the main
contribution to the gap. It can be interpreted as
the electrostatic energy of two uniformly charged rings of
charge $e$, radius $r_0$, at distance $d$ from each other, $l_B$
being the effective ring thickness. These rings correspond to an
electron and a hole uniformly spread along their cyclotron
orbits.

The contribution (\ref{capacitor})
is proportional to the magnetic field.
In Refs.~\onlinecite{Ashoori},~\onlinecite{Turner}
a gap linear
in the magnetic field was reported.
Our result for the gap is in agreement
with this data, however, numerically it is somewhat bigger.
It is natural, since the real charge
distribution during the
tunneling is more smeared than in our calculation,
because $l_B$ is not very small, and also because the  thickness $w$
of the  2D electron layer is comparable
to the distance between electrons.
Hence, our calculation gives
only the upper bound to the gap.

\subsection{Effect of a parallel field}

Let us consider the effect of a magnetic field parallel to the
layers. This is interesting because the field effect on the $I-V$
curve reveals the spatial coherence of the tunneling. The parallel
magnetic field does not affect the motion in the plane. However,
the parallel field flux, ``captured'' between the electron and
hole trajectories, will give an Aharonov--Bohm phase to the
tunneling amplitude. Thus, the tunneling current will oscillate
as the parallel field is swept.

To incorporate the parallel  field
in the formalism,
one may write its vector potential
$A_z = B_{\parallel} y$, $A_x = A_y =0$
($B_\parallel$ is assumed to be parallel to
the $x-$axis),
and then change the phase of $\psi_{R}$:
$\psi_{R}(x,y) \rightarrow
\psi_{R}(x,y) e^{ i eB_{\parallel}d y/\hbar c}$.
(Here   we restore $\hbar$ to show the characteristic
scale of $B_{\parallel}$.)
Hence, the amplitude $K$ acquires
an additional factor:
\be
\label{K-tilde}
\tilde K({\bf r},t)=\,
e^{-{ie\over\hbar c}B_\parallel d y}
K({\bf r},t)\ ,
\ee
which is just the Aharonov--Bohm phase mentioned above.
For the cyclotron orbit $y = R_c (\cos\theta_0 - \cos\theta_1)$, and
Eq.(\ref{current})
in the limit $eV \ll \hbar \omega_c$
yields (see Appendix B)
\be
\label{parallel-field}
I(V,B_\parallel) =
J_0^2\left(\frac{e d B_\parallel R_c}{\hbar c }\right)
I(V, B_\parallel = 0).
\ee
One may note that the tunneling current suppression by a factor
oscillating as a function of the parallel field is similar to the
magnetic field effect on the critical current in a wide
Josephson contact.

Note that the scale of spatial coherence $R_c$ is
much less than the scale of the charge spreading during
the tunneling, which is given by $a_0\sim 1/B^2$.

For an ideal two dimensional system, the current dependence on
$B_\parallel$ factors out because the parallel field does not
affect the motion in the plane. For real wells of finite width, the
factorization (\ref{parallel-field}) should still be a good
approximation at small $B_\perp$, when the cyclotron radius is
big compared to the well width. However, since the parallel
field will squeeze the states in the wells, and effectively
increase the barrier width, the tunneling rate may acquire an
additional non-oscillatory suppression factor.

\section{Conclusion}
The multi-dimensional  bosonization  technique is capable of dealing
with the long wavelength  excitations  in  a  Fermi  liquid.  We
applied this formalism to study tunneling into a Fermi liquid in
a magnetic field, where one has to account for  the  Fermi  liquid
response  to injection or removal of an electron. In the absence
of the field, there is a peak of the tunneling conductance  near
zero  bias.  In a  finite  field,  the  peak  is shifted to higher
voltage, and a tunneling gap is formed. We calculate the gap  at
the  field  weaker than the Quantum Hall field, and find that it
scales linearly with the field. We compare  these  results  with
the many-body calculation, and with experiments.

A particular advantage of the bosonization  method  is  that  it
gives  an  explicit  relation  with  the  classical  dynamics of
cyclotron motion, and with cyclotron orbits. We  find  that  the
tunneling is spatially coherent on the scale set by the cyclotron
radius,  and propose to probe this coherence by a magnetic field
parallel to the barrier. The effect of the parallel field on the
tunneling is due to an  Aharonov-Bohm  phase  in  the  tunneling
amplitude, which leads to oscillations of the current taken as a
function of the parallel field.

\appendix
\section{Tunneling current in Luttinger liquid}
We derive here Eq.(\ref{I-Luttinger}) from
Eqs.(\ref{current}) and (\ref{K-Luttinger}).
First, it is useful to integrate
$K(x,t)$ in Eq.(\ref{K-Luttinger}) before making an analytic
continuation.
By use of the identity
\be
\label{identity-a-1}
\int\limits_{-\infty}^{\infty}
\frac{dx}{\left(x^2 + a^2\right)^{\alpha}} =
\frac{\sqrt{\pi}}{|a|^{2\alpha-1}}
\frac{\Gamma\left(\alpha -
\frac{1}{2}\right)}{\Gamma\left(\alpha\right)}
\ee
one gets
\be
\label{K-Luttinger-integrated}
\int\limits_{-\infty}^{\infty} K(x,\tau)\,dx  = -
\frac{\alpha^{2(\beta-2)}}{2 \pi^{3/2} (v |\tau|)^{2\beta-3}}
\frac{(\beta-2)\,\Gamma\left(\beta-\frac{3}{2}\right)}{\Gamma(\beta)}
\ee
($x\Gamma(x)=\Gamma(x+1)$ was also used).
Now, let us continue this expression from the upper half-axes
to the upper half-plane and from the lower half-axes to the lower
half-plane. For $t>0$ it gives:
\begin{eqnarray}
\label{K-Luttinger-integrated-continued}
\int\limits_{-\infty}^{\infty}
\left(
K^{>}(x,t) - K^{<}(x,t)
\right) dx = \\
-i \frac{\alpha^{2(\beta-2)}}{\pi^{3/2} (vt)^{2\beta-3}}
\sin\frac{\pi}{2}(2\beta -3)
\frac{(\beta-2)\,\Gamma\left(\beta - \frac{3}{2}\right)}{\Gamma(\beta)}
\nonumber
\end{eqnarray}
Now, the integration over $t$ can performed by using the formula
\be
\int\limits_{0}^{\infty} \frac{e^{-i x}}{x^{\alpha}}  dx =
\frac{\pi e^{\frac{i \pi}{2}(\alpha - 1)}}{\sin\pi\alpha\Gamma(\alpha)}\
,
\ee
and the result is given by Eq.(\ref{I-Luttinger}).

\section{Effect of parallel field}

Here we derive Eq.(\ref{parallel-field}).
We will do it for free electrons, but a
generalization for a problem with
an interaction will be straightforward, as
 can be seen from Eq.(\ref{MainResult}) and the preceding
discussion.
First, let us integrate over $d{\bf  r}$ in
Eq.(\ref{current}) by noticing that
\begin{equation}
\label{change-of-variables}
r = 2 R_c \sin\frac{\theta_1 - \theta_0}{2}\,,
\qquad \varphi = \frac{\theta_0 + \theta_1}{2}\,,
\end{equation}
so that the  area element $rdrd\varphi$ can be
expressed in terms of $\theta = \theta_0 - \theta_1$ and
$\varphi$. Then one plugs
Eqs.(\ref{K-tilde}) and (\ref{K-continued})
 into Eq.(\ref{current}).
Integration over $\varphi$ produces a
zero-order Bessel function, and one arrives at
\begin{eqnarray}
I = \frac{2}{\pi}\frac{e|t_0|^2}{v_F \omega_c p_F} eV
\int\limits_{0}^{\infty} \cos{\frac{eV}{\omega_c}}e^{-\theta R_c/l_s}
\times\nonumber\\
J_{0} \left( \frac{2eB_\parallel d R_c}{\hbar c}
\sin\frac{\theta}{2}\right)
d\theta
\label{uzhas}
\end{eqnarray}
(here the interval of integration was extended to $(0,\infty)$, which
accounts for
the presence of the sum).
We make a Fourier expansion of $J_0(2a\sin\theta/2)$, by
using the identity
%
\be
\label{seria}
J_{0}(2a\sin\frac{\theta}{2})
=
\sum\limits_{n=-\infty}^{\infty}
e^{in\theta} J_{n}^{2}(a)\,,
\ee
and one gets
\be
\label{uzhas-2}
I = \frac{2}{\pi}\frac{e|t_0|^2}{l_s p_F} eV
\sum\limits_{-\infty}^{\infty}
\frac{J_{n}^2
      \left(\frac{eB_\parallel dR_c}{\hbar c}
      \right)
}{(eV - n \omega_c)^2 + \frac{v_F^2}{l_s^2}}\,.
\ee
Since we are dealing with the limit $eV \ll \hbar \omega_c$,
only the $0-$th term is essential, and one arrives at
Eq.(\ref{parallel-field}). Finally,
Eq.(\ref{parallel-field})
 predicts the total
suppression of the tunneling current at certain values of
$B_\parallel$, given by the roots of the Bessel function $J_0$.
However, near these points the current $I(B_\parallel)$ vanishes.
This is an artifact of the approximation. If one keeps the higher
order terms in Eq.(\ref{uzhas-2}), the minimal values of the current
become positive.


\begin{references}
\bibitem{Gornik}
J. Smoliner, E. Gornik, and G. Weimann,
{\it Appl. Phys. Lett.} {\bf 52}, 2136 (1988);\\
W. Demmerle, et al.,
{\it Phys. Rev. B} {\bf 44}, 3090 (1991)
\bibitem{Eisenstein'91}
J. P. Eisenstein, T. J. Gramilla, L. N. Pfeiffer, and K. W. West,
{\it  Phys. Rev. B}, {\bf 44}, 6511 (1991)
\bibitem{Eisenstein'91a}
J. P. Eisenstein, L. N. Pfeiffer, and K. W. West,
{\it Appl. Phys. Lett.}, {\bf 58}(14), 1497 (1991)
\bibitem{Eisenstein'92}
J. P. Eisenstein, L. N. Pfeiffer, K. W. West,
{\it  Phys. Rev. Lett.}, {\bf 69}, 3804 (1992)
\bibitem{Eisenstein'94}
J. P. Eisenstein, L. N. Pfeiffer, K. W. West,
{\it Surf. Sci.}, {\bf 305}, 393 (1994)
\bibitem{Eisenstein'95}
J. P. Eisenstein, L. N. Pfeiffer, K. W. West,
{\it Phys. Rev. Lett.}, {\bf 74}, 1419 (1995)
\bibitem{Ashoori}
R. C. Ashoori, J. A. Lebens, N. P. Bigelow, and
R. H. Silsbee, {\it Phys. Rev. Lett.}, {\bf 64}, 681(1990);
{\it Phys. Rev. B}, {\bf 48}, 4616 (1993).
\bibitem{Brown}
K. M. Brown, N. Turner, J. T. Nicholls, et. al.,
{\it Phys. Rev. B}, {\bf 50}, 15 465 (1995)
\bibitem{Turner}
N. Turner, J. T. Nicholls, K. M. Brown,  et. al.,
``Tunneling Between Two-Dimensional
Electron Gases in a Weak Magnetic Field,'' preprint, cond-mat/9503040
\bibitem{controvercy}
Let us mention that the behaviour  of
the gap in the intermediate field range, seen by different
groups\cite{Brown,Eisenstein'92}, is controversial.
\bibitem{ZhengMacDonald}
L. Zheng, and A. H. MacDonald,
{\it Phys. Rev. B}, {\bf 47}, p. 10619 (1993)
\bibitem{Halperin}
S. He, P. M. Platzman, B. I. Halperin, {\it Phys. Rev. Lett.}, {\bf 71},
777 (1993);\\
Y. Hatsugai, P. A. Bares, and X.G. Wen, {\it Phys. Rev. Lett.},{\bf 71},
424 (1993);
\bibitem{HighFieldTheory}
S. R. E. Yang and A. H. MacDonald, {\it Phys. Rev. Lett.},{\bf 70}, 4110
(1993);\\
P. Johansson and J. M. Kinaret, {\it Phys. Rev. Lett.}{\bf 71}, 1435
(1993);\\
S. R. Renn and B. W. Roberts, {\it Phys. Rev.}{\bf B 50}, 7626 (1994);
\bibitem{Efros}
A. L. Efros and F. G. Pikus, {\it Phys. Rev.}{\bf B 48}, 14694 (1993);\\
A. L. Efros and F. G. Pikus, Proceedings of the International Conference
on High Magnetic Fields in Semiconductor Physics, Boston, 1994
\bibitem{HighFieldExcitonic}
C. M. Varma, A. I. Larkin, and E. Abrahams, {\it Phys. Rev.}{\bf B 49},
13999 (1994);
\bibitem{Glazman}
I. L. Aleiner, H. U. Baranger, and L. I. Glazman,
{\it Phys. Rev. Lett.},{\bf 74}, 3435 (1995)
\bibitem{Aleiner}
I. L. Aleiner and L. I. Glazman,
preprint, cond-mat/9505026
\bibitem{Mahan}
G. D. Mahan, {\it Many-Particle Physics}, Sec. 9.3
 (Plenum Press, New York, 1981)
\bibitem{Negele}
J. W. Negele
{\it Rev. Mod. Phys.} {\bf 54}, 913 (1982);\\
S. Levit, J. W. Negele, and Z. Paltiel,
{\it Phys. Rev. C} {\bf 22}, 1979 (1980)
\bibitem{Luther}  A. Luther, Phys.Rev. B, {\bf 19}, 320 (1979).
\bibitem{Jackiw} R. Floreanini and R. Jackiw,
{\it Phys. Rev. Lett}, {\bf 59}, 1873 (1987)
\bibitem{Haldane-1D} F. D. M. Haldane, {\it J. Phys. C}, {\bf 14}, 2585
(1981)
\bibitem{Haldane} F. D. M. Haldane, "Luttinger theorem and Bosonization
of the Fermi Surface", {\it Proceedings of the International School of
Physics
"Enrico Fermi"}, {\bf 121}, Varenna 1992, edited by R. Schrieffer
and R. A. Broglia (North-Holland, New York, NY 1994).
\bibitem{Fradkin} A. H. Castro Neto, E. H. Fradkin,
{\it Phys. Rev. Lett.}, {\bf 72 }, 1393 (1994);
{\it Phys. Rev. B}, {\bf 49}, 10 877 (1994);
"Exact solution of Landau fixed Point", preprint, cond-mat/9310046.
\bibitem{Houghton} A. Houghton and J. B. Marston,
{\it Phys. Rev B}, {\bf 48}, 7790 (1993).
\bibitem{Kadanoff}
L. Kadanoff and G. Baym, {\it Quantum Statistical Mechanics}
(Addison Wesley, 1991)
\bibitem{bounce}
J. S. Langer,
{\it Ann. Phys. (N.Y.)} {\bf 41}, 108 (1967);\\
S. Coleman,
{\it Phys. Rev. D} {\bf 15}, 2929 (1977)
\bibitem{Luttinger-Green}
V. J. Emery,
in: {\it Highly Conducting One-Dimensional Solids},
eds. J. T. Devreese, R. P. Evrard, and V. E. van Doren (Plenum, 1979);\\
Ref.~\onlinecite{Mahan}, Sec. 4.4
\bibitem{AGD} A. A. Abrikosov, L.~P.~Gorkov,
I.~E.~Dzyaloshnskii, {\it  Methods of Quantum Field Theory in
Statistical Physics}, (Dover, New York, 1975)
\bibitem{QFT} see any text on quantum field theory, where
scattering amplitudes are regularized by a finite volume.
\bibitem{Gutzwiller} M. C. Gutzwiller,
{\it Chaos in Classical and Quantum Mechanics}, Sec. 12.5
(Springer-Verlag, New York, 1990)
\bibitem{LS1} L. S. Levitov, A. V. Shytov, ``Coulomb zero-bias
anomaly: A semiclassical calculation,'' preprint, cond-mat/9501130.
\bibitem{LS2} L. S. Levitov, A. V. Shytov,
``Spatial coherence of tunneling in double wells,''
preprint, cond-mat/9507058
\end{references}
\end{document}